\documentclass[a4paper,fleqn]{cas-sc}

\usepackage[numbers]{natbib}
\usepackage{color}

\newcommand{\ve}[1]{{\text{\bf #1}}} 
\newcommand{\vk}{\ve k}
\newcommand{\vp}{\ve p}

\newcommand{\vx}{\ve x}

\begin{document}
\let\WriteBookmarks\relax
\def\floatpagepagefraction{1}
\def\textpagefraction{.001}
\shorttitle{Effects of Dark Energy anisotropic stress on the matter power spectrum}
\shortauthors{Garcia-Arroyo et~al.}

\title [mode = title]{Effects of Dark Energy anisotropic stress on the matter power spectrum}                      

\author[1]{Gabriela~Garcia-Arroyo}[orcid=0000-0002-0599-7036]
\cormark[1]
\ead{gaby.7646@gmail.com}

\author[2]{Jorge~L.~Cervantes-Cota}[orcid=0000-0002-3057-6786]
\ead{jorge.cervantes@inin.gob.mx}

\author[1,3,4]{Ulises~Nucamendi}[orcid=0000-0002-8995-7356]
\ead{unucamendi@gmail.com}

\author[2,5]{Alejandro~Aviles}[orcid=0000-0001-5998-3986]
\ead{avilescervantes@gmail.com}

\address[1]{Instituto de F\'isica y Matem\'aticas,
             Universidad Michoacana de San Nicol\'as de Hidalgo,
             Edificio C-3, Ciudad Universitaria, CP. 58040
             Morelia, Michoac\'an, M\'exico.}

\address[2]{Departamento de F\'isica, Instituto Nacional de Investigaciones Nucleares, A.P. 18-1027, Col. Escand\'on, CDMX, 11801, M\'exico.}  

\address[3]{Departamento de F\'isica, Cinvestav, Avenida Instituto Politecnico Nacional 2508, San Pedro Zacatenco, 07360 Gustavo A. Madero, Ciudad de Mexico, Mexico.}

\address[4]{Mesoamerican Centre for Theoretical Physics, Universidad Autonoma de Chiapas, Carretera Zapata Km. 4, Real del Bosque (Teran), 29040, Tuxtla Gutierrez, Chiapas, Mexico.}             

\address[5]{Consejo Nacional de Ciencia y Tecnolog\'ia,
Av. Insurgentes Sur 1582, Colonia Cr\'edito Constructor, Del. Benito Ju\'arez, 03940, 
CDMX, M\'exico.}

\cortext[cor1]{Corresponding author}

\begin{abstract}
We study the effects of dark energy (DE) anisotropic stress on features of the matter power spectrum (PS). We employ the Parametrized Post-Friedmannian (PPF) formalism to emulate an effective DE, and model its anisotropic stress properties through a two-parameter equation that governs its overall amplitude ($g_0$) and transition scale ($c_g$). For the background cosmology, we consider different equations of state to model DE including a constant $w_0$ parameter, and models that provide thawing (CPL) and freezing (nCPL) behaviors. We first constrain these parameters by using the Pantheon, BAO, $H_0$ and CMB Planck data. Then, we analyze the role played by these parameters in the linear PS. In order for the anisotropic stress not to provoke deviations larger than $10\%$ and $5\%$ with respect to the $\Lambda$CDM PS at  $k \sim 0.01 \,h/\text{Mpc}$, the parameters have to be in the range  $-0.30< g_0 < 0.32$, $0 \leq c_g^2 < 0.01$ and $-0.15 < g_0 < 0.16$, $0 \leq c_g^2 < 0.01$, respectively.  
Additionally, we compute the leading nonlinear corrections to the PS using standard perturbation theory in real and redshift space, showing that the differences with respect to the $\Lambda$CDM are enhanced, especially for the quadrupole and hexadecapole RSD multipoles.
\end{abstract}

\maketitle
\section{Introduction}
\label{sec:intro}

The discovery of the accelerated expansion of the Universe implied the existence of dark energy (DE), that has been extensively confirmed by a two-decade variety of experiments, initially employing Supernovae type Ia \cite{Riess:1998cb,Perlmutter:1998np}, then using anisotropies in the CMB from WMAP and Planck data \cite{2016A&A...594A..13P}, distance measurements of different  tracers \cite{Stern:2009ep}, and clustering of large galaxy surveys, among other probes \cite{Blake:2011en,Abbott:2017wau,Ata:2017dya}. However, little is known of the fundamental properties of DE, apart from being a `fluid' that possess negative pressure. In the most successful model a cosmological constant, $\Lambda$, is capable to fit the observations, albeit current tensions exist among a few parameters when measured with  different probes \cite{Riess:2019cxk}.

The effects of DE have been widely studied in the context of background cosmological dynamics; most of the work has been devoted to test different equations of state (EoS) for DE to understand the dynamics of the Hubble expansion flow. However, in comparison its perturbative effects are less explored, partially because we expect little deviations at perturbative level, but also because we have no clues on its fundamental origin.
 One can, for example, treat DE as a barotropic fluid, hence, its sound speed depends only on background quantities, or to consider it as a non-adiabatic fluid to account for its linear effects for which additional hypotheses have to be made about the fluid's speed of sound \cite{Amendola:2015ksp}. There are many works that study the effect of DE speed of sound in the perturbative dynamics, 
initially done by   \cite{Avelino:2002fj,Sandvik:2002jz,Balakin:2003tk, PhysRevD.69.083503}. It turns out that the effects of DE clustering result to be small, especially if the DE EoS is close  to $-1$, as demanded by observations, and then they are difficult to discern with late-Universe measurements \cite{Weller:2003hw,Huterer_2017}. But, in fact, varying the DE sound speed can induce deviations of up 2$\%$ in the matter power spectrum (PS) \cite{dePutter:2010vy,Vagnozzi:2019kvw}, that should be important in view of the expected constraints from upcoming galaxy surveys, such as DESI \cite{Aghamousa:2016zmz}. 

Another possibility is to consider DE anisotropic stress. A homogenous and isotropic symmetric background  metric forbids it, but it can be introduced at the perturbed level \cite{Hu:1998tj}.  Anisotropic stress can also mimic modified gravity (MG) at linear order \cite{Kunz_2007,Song:2010,Arjona_2019,Arjona_2019b}, since it introduces at least a new parameter, and together with DE EoS and sound speed, it yields a modified growth of structures in the Universe. In fact, DE stress generates similar outcomes as those of varying the sound speed of DE, but the detailed  behavior depends on the signs of the EoS and stress parameter \cite{Koivisto:2005mm}.  From theoretical grounds, one expects DE anisotropic stress to affect the evolution of the metric potentials and this provokes CMB temperature anisotropies at low-multipoles, to be affected through the Integrated Sachs-Wolfe (ISW)  effect. In Refs.~\cite{Hu:1998kj,Koivisto:2005mm,PhysRevD.77.103524,Ichiki_2007,Calabrese_2011,Chang_2014} DE anisotropic stress was analyzed to prove this conclusion using CMB data available at that time,  but due to the cosmic variance, CMB constraints are still broad.  However, DE stress should affect also matter clustering at large scales. Effects of anisotropic stress on the matter power spectrum (PS) and on the growth function have been studied in several works \cite{Hu:1998kj,Koivisto:2005mm,Mota:2007sz,Pogosian_2010,Sapone_2012,Cardona_2014,PhysRevD.90.103528}, showing that shear viscosity has an effect on very large scales, but one the other hand it does not change much other cosmological parameter values; for instance, for this latter reason we do not expect that DE shear terms alone can alleviate the current tension in the Hubble constant; see however \cite{Yang_2018} in which it is proven that adding anisotropic shear to interacting models helps to increase the Hubble constant to release the tension for phantom DE.

In the literature there is a number of works considering different aspects of imperfect fluids, e.g. in connection to second order perturbation in $\Lambda$CDM \cite{Ballesteros_2012}, or related to generalized scalar fields \cite{Sawicki_2013,Naruko_2019}.  Also, based on MG, efforts have been put forward to understand how the gravitational effects of the fifth-force (that generates an effective shear term) influence the observables at cosmological scales, changing the clustering properties \cite{Huterer_2010,Baker_2014}. Our motivation here, linked to these latter works, is  to analyze the anisotropic stress effects on CMB and matter PS since the level of accuracy of future LSS galaxy surveys and probes shall demand detailed understanding of the clustering properties of the matter field. In this way, being able to constrain an hypothetical anisotropic shear, stemming either from DE or MG. Ways to carry out this comparison are discussed e.g. in Refs. \cite{Motta_2013,Amendola_2014,Pinho_2018}.   Recently, analysis of recent probes hints for non-zero anisotropic stress \cite{arjona2020hints}, that also encourages us to further analyze its clustering properties.      

In the present work, we use the Parametrized Post-Friedmannian (PPF) approach \cite{Hu:2007pj,PhysRevD.77.103524,PhysRevD.78.087303}, though originally motivated to emulate MG models, they naturally introduce an effective DE anisotropic stress term. In this respect, the formalism serves to phenomenologically introduce either DE stress or  effective MG stress terms. Viewed as effective DE, we consider specific equations of state and fix the DE speed of sound, to concentrate our analysis on the effects of the anisotropic stress. We analyze the constraints from CMB power spectra and, especially, look for deviations in the PS. Interestingly, we find that DE anisotropic stress is allowed by Planck CMB data, as in Refs.  \cite{Koivisto:2005mm,PhysRevD.77.103524}, but the linear and nonlinear PS impose tighter constraints to it. We consider different DE EoS, firstly $w=-1$ that emulates $\Lambda$ at background level,  then constant $w_0$, and finally, thawing and freezing models, to find out their effects in combination with stress parameters. 

The structure of this paper is the following: In Section \ref{sec:background} we motivate DE EoS  parametrizations chosen, and in Section \ref{linear-perturbs} we introduce the DE perturbation theory with anisotropic stress, where a specific anisotropic stress phenomenology is adopted.  Section 
\ref{results} shows our results employing different EoS, and Section \ref{nonlinear_PS} shows the theory and results for nonlinear perturbation theory to 1-loop. Finally, Section \ref{conclusions} concludes.

\section{Dark energy Equation of State}
\label{sec:background}

Beyond a cosmological constant, the accelerated expansion of the Universe can be driven by  a dynamical DE component whose EoS is commonly parametrized by a time dependent function 
\begin{equation}
P_{de}=w(z)\rho_{de},    
\end{equation}
 where the EoS parameter, $w(z)$, can be chosen with different purposes; as for example, it can mimic quintessence and phantom fields \cite{Macorra_2016,Garc_a_Garc_a_2020}. In general, EoS parameterizations $w(z)$ can be classified into two broad categories: thawing and freezing behaviors \cite{Caldwell:2005tm,PhysLetB686}.  In the first case the scalar field is frozen at early times where the kinetic energy is negligible and $w\sim -1$, then $w(z)$ evolves generically as a monotonic, convex, decreasing function to reach asymptotically, at late times, some $w\geq -1$.
In the second case, in freezing-tracker models, the scalar field rolls down to the minimum of its potential at the beginning of the Universe, but starts to slow down and stops when it comes to dominate the dynamics; in this case the $w(z)$ function is generically a  monotonic, concave, increasing function at higher $z$ which at late times tends to $w\sim -1$. Several DE parametrizations have been proposed in the literature, some seem to favor thawing models \cite{Planck2016,Hara:2017ekj}, but  Ref.~\cite{PhysRevD.93.103503} exhibits that  freezing models fit better. This latter reference proposes a generalization of the CPL EoS \cite{Chevallier:2000qy, Linder:2002et}, called nCPL,  
\begin{equation}\label{eq:ncpl}
w(z)= w_0 + w_a\left(\frac{z}{1+z}\right)^{n} ,
\end{equation}
such that $n=1$ reduces to the standard CPL (suitable for thawing models) while for larger values of $n$ it can produce freezing behavior. 
Since our goal is to understand how different EoS behaviors influence CMB anisotropies and the PS, especially in combination with anisotropic stress, we will consider the nCPL parametrization with $n=1$ and $n=7$, corresponding to thawing and freezing behaviors, respectively.  We will also consider
 models with  $w$ constant. 
For any $n>0$, the nCPL EoS parametrization is $w_0$ at $z=0$ and goes to $w_0+w_a$ at high redshifts. A requirement to  achieve a thawing behavior is that the function should be decreasing as $z$ grows and so $w_a$ must be negative, and to get a freezing evolution $w_a$ must be positive. 
 
The energy density for nCPL evolves as
\begin{equation}\label{eq:dedensity}
\rho_{de}(a) = \left \{
\begin{array}{l}
\rho_{de}^{0}a^{-3(1+w_{0}+w_{a})}\exp{\left[-3w_{a}(1-a) \right]} \quad (n=1), \\
{}\\ {}\\
\rho_{de}^{0}a^{-3(1+w_{0}+w_{a})}\exp\left[-3w_a\left(\frac{363}{140}-7a+ \frac{21}{2}a^2-\frac{35}{3}a^3+\frac{35}{4}a^{4}-\frac{21}{5}a^5+\frac{7}{6}a^{6}-\frac{1}{7}a^{7} \right)\right] \quad (n=7), \\{}
\end{array} 
\right.
\end{equation}
which together with the other matter components it determines the background history, $H(z)$, through the Friedmann equation.
\section{Dark energy fluctuations}
\label{linear-perturbs}
We consider a perturbed metric around a  Friedmann-Lemaitre-Robertson-Walker (FLRW) spacetime in Newtonian, longitudinal gauge,
\begin{equation}\label{eq:metric}
    ds^2=a^{2}(\tau)\left[-\left(1+ 2 \Psi \right)d\tau ^2 + (1+2\Phi)dx^{i}dx_{i}\right],
\end{equation}
where  $\Psi$ and $\Phi$ are the gauge invariant scalar potentials \cite{PhysRevD.22.1882,Kodama}. 
The components of the energy momentum tensor are
\begin{align}
T^{0}{}_{0} &=-\left(\rho + \delta \rho\right), \nonumber\\ 
T^{0}{}_{k} &=\left(\rho + P\right)v_k, \nonumber \\
T^{k}{}_{l} &= \left(P+\delta P \right)\delta^{k}{}_{l}+ P \Pi ^{k}{}_{l},
\end{align}
where $\rho$ and $P$ are the energy density and pressure at the background, $\delta \rho$ and $\delta P$ their pertubations, 
and $\Pi^{k}{}_{l}$ are the anisotropic stress components. 
Since we are dealing with scalar perturbations, it is useful to work with the velocity divergence $\theta$ and the scalar anisotropic stress $\Pi$ defined as
\begin{align}
 \theta &= i k^i v_i, \\
 \Pi &= -\frac{3}{2} \left( \frac{k_i k_j}{k^2} -\frac{1}{3} \delta_{ij} \right) \Pi^{ij}.
\end{align}
From Einstein's field equations one  obtains 
\begin{align}
k^{2}\Phi &= 4\pi G a^{2} \sum_{{\rm a}}\left(\delta \rho_{{\rm a}} + 3Ha(\rho_{{\rm a}} + P_{{\rm a}})\frac{\theta_{{\rm a}}}{k^{2}} \right) \, ,  \label{eq:Poisson} \\ 
    k^{2} \left( \Phi + \Psi \right)& =  -8\pi G a^2 \sum_{\rm a} P_{\rm a} \Pi_{\rm a} \, ,\label{eq:Shear}
\end{align}
where the sums run over all energy components.  In the absence of anisotropic stresses both gravitational potentials are equal (up to a minus sign). At early times the difference 
in the two gravitational potentials is sourced by the second moment of the phase-space distribution function of radiation components. 
However at late times, well after decoupling and during the matter dominated phase, this is negligible and one can safely set $\Psi=-\Phi$ to obtain
the standard growth of matter linear perturbations $\delta_m \propto a$. This is a key property of cold dark matter (CDM), allowing its perturbations to grow at the same rate for all scales well below the Hubble horizon
during the matter dominated phase. At later times, once DE starts to become important, the growth of large scales structures is halted because the expansion becomes very
fast and the pace of matter aggregation is reduced,  even frozen for a de Sitter expansion.  
The details of how this process occurs depend on the very nature of DE. 
Matter components source the gravitational potential $\Phi$ through the Poisson equation,  Eq.~(\ref{eq:Poisson}), however the trajectories of non-relativistic CDM particles respond to the gravitational potential $\Psi$ through the
geodesic equation, which in the Newtonian limit is $\ddot{\vx} = -\nabla \Psi$. 
Hence, even if probes of the Universe's expansion indicate that DE should very close to a cosmological constant with $P \approx - \rho $, the growth of perturbations can be very different in the presence of
the anisotropic stress $\Pi_{de}$. But note that this quantity is not accessible from background observations, and by taking a posture of complete ignorance about what DE is, it is natural to incorporate the stress in a perturbative analysis,
on the same footing as one introduces the EoS and the speed of sound. The anisotropic stress should be small at early times, before decoupling, in order to not spoil the CMB anisotropies, tightly constraining models and leaving  room
to affect the CMB only through the ISW effect. Hence, it is expected that effects of a DE stress will be more feasible to be detected through CDM late time clustering probes, in particular the matter PS.

From the conservation of the energy-momentum tensor we get the continuity and Euler equations, for non-interacting fluids these reduce to
\begin{equation}\label{eq:delta}
\delta {\rho '} + 3(\delta \rho + \delta P) = - (\rho + P )\left(3{\Phi '}+\frac{\theta}{aH}\right) \, ,
\end{equation}
\begin{equation}\label{eq:velocidad}
 (\rho + P) \theta' =  \frac{k^2}{aH} \big( \Psi (\rho +P) + \delta P \big) -(\rho'+ P') \theta -4 (\rho + P) \theta-\frac{2}{3aH} k^{2} P \Pi ,
\end{equation}
where we use derivatives with respect to $\ln(a)$, denoted by a prime.
At the background level adiabaticity is guaranteed by the continuity equation, however, when fluctuations are considered, the energy density of components and their EoS do not completely specify  their pressure. In a general description,
for non-interacting components one has the relation \cite{PhysRevD.79.023502,PhysRevD.97.103514,PhysRevD.69.083503}
\begin{equation}\label{eq:pressure}
\delta P= c_{s}^{2} \delta \rho+3aH(\rho + P)(c_{s}^{2}-c_{a}^2)\frac{\theta}{k^{2}} ,
\end{equation}
with  $c_{a}$ the adiabatic sound speed and $c_{s}$ the speed of sound in the fluid's rest frame,
\begin{equation}
c_{a}^2\equiv \frac{{P'}}{{\rho'}}, \quad c_{s}^2 \equiv \frac{\delta P^{rest}}{\delta \rho^{rest}} ,
\end{equation}
where 
\begin{equation}\label{eq:delta_rest}
\delta \rho^{rest}= \delta \rho+3Ha(\rho + P) \frac{\theta}{k^{2}} \, 
\end{equation}
is the gauge invariant rest-frame density perturbation \cite{PhysRevD.22.1882}.

\subsection{Dark energy anisotropic stress} \label{DE-ani_stress}

There are different approaches to implement  
the evolution of DE anisotropic stress to solve the system \eqref{eq:Poisson}-\eqref{eq:velocidad}. Some works \cite{Koivisto:2005mm,Hu:1998kj,Calabrese_2011,Chang_2014,Sapone_2012,PhysRevD.90.103528,Aviles:2014mua} assume  
that anisotropic stress is sourced by the amplitude of the velocity shear tensor $\partial^{i} v^{j}_{de}$, and are motivated by the fact that it should be gauge invariant;
so they demand to fulfill  a continuity-like equation stemming from a Boltzmann hierarchy, 
but invoking an effective viscosity parameter as a source. This approach washes out DE fluctuations for non-phantom EoS \cite{Mota:2007sz}, making them even more difficult to detect when compared to other approaches, as those 
motivated by MG  \cite{Cardona_2014} or modified growth \cite{Song:2010}, where effects inside the horizon are also expected. Both approaches are valid and are motivated by different physics. The specific model will then determine the effects on observables. Our motivation is to explore effects on scales that will be available through the next generation of LSS measurements.

In this work, we will use the PPF prescription presented in Refs.  \cite{Hu:2007pj,PhysRevD.77.103524,PhysRevD.78.087303}, originally motivated to parametrize MG models that naturally introduce an effective anisotropic stress term. In this view, DE is a phenomenon of a geometric theory.  This approach studies perturbation modes larger and smaller than the horizon and, therefore,  
it needs to impose two conditions on the field equations in order to preserve covariant conservation laws for the fluids. The first condition requires that the curvature in the comoving gauge is only changed by the effective DE at 
second order in $k_H \,(\equiv k/aH )$. The second condition is that the metric potential satisfies a Poisson-like equation in the quasi-static limit. Following this formalism, one introduces the effective DE shear in Eq.~\eqref{eq:Shear} through
\begin{equation} \label{phi+}
  \Phi_{+}\equiv g(a,k)\Phi_{-}-\frac{4\pi G a^2}{k^2} P_{T}\Pi_{T},  
\end{equation}
where 
$\Phi_{-} \equiv \frac{\Phi-\Psi}{2},\quad \Phi_{+} \equiv \frac{\Phi + \Psi}{2}$, and the subindex $T$ is for the sum of total matter components,  that in our case excludes DE.  
Equation (\ref{phi+}) defines the  function $g(a,k)$, which relates the two metric potentials and encodes the information of effective DE anisotropic stress as
\begin{equation}\label{eq:de_stress}
    P_{de}\Pi_{de}= -g \frac{k^{2}\Phi_{-}}{4 \pi G a^{2}} ,
\end{equation}
thus, instead of using $\Pi_{de}$, one can work with $g$. In the absence of stresses other than DE, as it happens at late times, $g$ is related to the more commonly used slip parameter $\gamma \equiv -\Phi/\Psi$ as
\begin{equation}
g = \frac{\gamma-1}{\gamma+1},    
\end{equation}
reducing to DE stress-free models when $\gamma=1$. Below we will choose specific parametrizations for $g$ that are negligible at early times, when the stresses from radiation components are important. Hence, the stresses due to DE and other components are essentially not coupled, and we can think of this PPF formalism as a DE parametrization, instead of MG.  

Using Eqs.~(\ref{eq:Poisson}) and (\ref{eq:Shear}) we obtain the constraint
\begin{equation}\label{eq:mod_poisson}
  k^{2}\Phi_{-}=4\pi G a^2 \left(\delta \rho_T^{rest} + \delta \rho_{de}^{rest}   +P_{de}\Pi_{de} +P_{T}\Pi _{T} \right),  
\end{equation}
which reduces to the Poisson equation, Eq.~\eqref{eq:Poisson}, in the absence of anisotropic stresses.

The PPF formalism further introduces a couple of functions to encompass physical conditions in the limits of modes much larger and much smaller than the Hubble horizon.  At large scales, one has that 
\begin{equation} \label{f_zeta}
 \lim_{k_{H}\ll 1} \frac{4 \pi G}{H^2} (\rho_{de} + P_{de}) \frac{\theta_{de} - \theta_{T}}{k_H} = - \frac{1}{3} k_H \theta_{T} f_{\zeta} (a), \end{equation}
so that effective DE is parametrized by $f_{\zeta}$ at large scales.  

 In the opposite limit, for modes well inside the horizon, DE is smooth compared with matter and the metric potential satisfies 
\begin{equation}\label{eq:quasistatic}
    \lim_{k_H \gg1}\Phi_{-}=\frac{4\pi G a^2}{k^2}\frac{\delta \rho_T^{rest}+P_{T}\Pi _{T}}{(1+f_G)} , 
\end{equation}
which has the form of a  Poisson equation for the potential $\Phi_-$ with Newton's constant rescaled by $(1+f_G)^{-1}$ and it is sourced by anisotropic stresses of other components, different from  DE, that can be safely neglected at late times.

In addition, one introduces  the function $\Gamma$, which encodes the  DE contributions
and allows to express the source of an effective potential $\Phi_{-} + \Gamma$ in terms of matter variables only as 
\begin{align}\label{eq:poisson_gamma}
k^{2}(\Phi_{-}+\Gamma) &= 4\pi G a^2 \Big(\delta \rho_T^{rest} +P_{T} \Pi _{T} \Big).
\end{align}
A comparison between Eqs.~\eqref{eq:mod_poisson} and \eqref{eq:poisson_gamma} relates function $\Gamma$ with DE quantities
\begin{equation}\label{eq:Gamma}
\Gamma =-\frac{4\pi G a^2}{k^2}\Big(\delta \rho_{de}^{rest} +P_{de}\Pi_{de}\Big) .
\end{equation}
$\Gamma$ should fulfill the two above mentioned limits, hence the PPF formalism constructs the equation of motion \cite{Hu:2007pj,PhysRevD.77.103524}
\begin{equation}\label{eq:deGamma}
\left(1+c_{\Gamma}^{2}k_{H}^{2}\right)\left[\Gamma'+\Gamma+c_{\Gamma}^{2}k_{H}^{2}(\Gamma-f_{G}\Phi_{-})\right] = S \, ,
\end{equation}
where $c_{\Gamma}$ is  a constant that modulates the transition scale between the two limits, and
\begin{align}
S = \frac{g'-2g}{g+1}\Phi_{-} + \frac{4\pi G}{(g+1)k^{2} } 
\Big( g [(P_{T}\Pi_{T})'+  P_{T}\Pi_{T}] 
- [ (g+f_{\zeta}+g f_{\zeta})(\rho_T+P_T)  -(\rho_{de}+P_{de})] \frac{\theta_T}{H} \Big) ,
\end{align}
such that Eq.~\eqref{eq:deGamma} satisfies both limits
\begin{eqnarray}
    \lim_{k_{H}\ll 1} \Gamma'=-\Gamma + S  \, ,
&&
    \lim_{k_{H}\gg 1}\Gamma = f_{G}\Phi_{-} \, .
\end{eqnarray}
These are all the required equations and conditions for the DE PPF prescription. Once functions $g(a,k)$, $f_{\zeta}(a)$ and $f_G(a)$ are given, it is possible to solve for $\Gamma$ to  
finally recover the effective DE quantities. 

In terms of these functions the rest frame DE density perturbation is
\begin{equation}\label{eq:de_rest}
\delta \rho_{de}^{rest}=-\frac{(g+1) k^{2} \Gamma}{ 4\pi G a^2} +  g \delta \rho_{T}^{rest},
\end{equation}
and $\Phi_{-}$ can be obtained through
\begin{equation}\label{eq:isw}
  k^2\Phi_{-}= \frac{4\pi G a^2 (\delta \rho_{T}^{rest}+\delta \rho_{de}^{rest}+ P_{T}\Pi _{T})}{1+g} .
\end{equation}
The potential $\Phi_{-}$ is associated to the ISW effect, and we will obtain that for a positive, increasing function of time $g$, it will cause an increase of the low-$\ell$ CMB anisotropies. 

The DE anisotropic stress contribution can be written as:
 
 \begin{equation} \label{eq:PdePIde}
 P_{de}\Pi_{de}=-\frac{g}{1+g}\left[\delta \rho_{T}^{rest}+\delta \rho_{de}^{rest}+P_{T}\Pi _{T}\right] .
 \end{equation}
 By construction, in the PPF formalism, DE anisotropic stress  receives contributions of all energy density perturbations and the different  components are multiplied by the same factor. In contrast, other approaches \cite{Cardona_2014, 10.1093/mnras/sty2789} consider 
 different weights to the source terms of Eq. (\ref{eq:PdePIde}) aiming at suppressing shear terms at large scales.   

\subsection{Dark energy stress phenomenology}
To solve the equations in the PPF formalism, the functions $f_{\zeta}(a)$, $f_G(a)$ and $g(a,k)$, and the constants $c_{s}^{2}$ and $c_{\Gamma}^2$ should  be specified.
We know that DE becomes important and not negligible at large scales,  this is the limit $k_{H}\ll 1$, where we can expect $g(a,k) \neq 0$ and at least $\mathcal{O}(k_{H}^{-2})$, also we expect $\Gamma \neq 0$. Motivated by MG models,

 we set $f_{\zeta}=0.4 \, g_{SH}$, though its exact choice is rarely important for observable quantities, see \cite{Hu:2007pj,PhysRevD.77.103524}. Varying the DE sound speed ($c_s$) it is known to provoke variations of up 2$\%$ in the matter power spectrum \cite{dePutter:2010vy}, but we fix it a constant here, $c_s^2=1$, to concentrate our analysis on the anisotropic stress. And we further assume  $c_{\Gamma}= 0.4 \,c_s$, inspired to match the evolution of scalar fields,   following Ref. \cite{PhysRevD.78.087303}.
In the opposite limit, $k_{H}\gg 1$, we want to have an effective DE that emulates $\Lambda$CDM, thus we will fix $f_G=0$.

It is necessary to specify the anisotropic stress function $g(a,k)$, and its election  must obtain deviations when and where it is desired to test them. As explained, large modes affect the ISW signal, and we want to evaluate how much the cosmic variance allows deviations from  $\Lambda$CDM to later analyze 
the effect in the matter PS. Also, we would like to observe effects at different scales. In this work we use the function $g(a,k)$ proposed in \cite{PhysRevD.77.103524},
\begin{equation}\label{eq:gak}
    g(a,k)=\frac{g_{SH}(a)}{1+(c_{g}k_{H})^{2}},
\end{equation}
which becomes $k$-independent for $k_{H} \ll 1$ and goes quickly to zero at small scales. The constant $c_g$ is a transition parameter that determines the modes that plays a role with respect to the Hubble horizon.  On the other hand, DE should start to be important at late times when it becomes to dominate over  the  other matter components, so it seems natural to think of a dependency on the ratio  among densities which grows with $a$, motivating the time dependence form of $g_{SH}$ as
\begin{equation} \label{gSH}
    g_{SH}(a)=g_{0}\left(\frac{\rho_{de}(a)}{\rho_{T}(a)}\frac{\Omega_{T}^{0}}{\Omega_{de}^{0}}\right)^{1/2} .
\end{equation}
Hence, the function $g$ introduces two free parameters, its amplitude $g_0$ and the transition scale  $c_g$ along $k$-modes. Different combinations of the two anisotropic stress parameters can achieve similar effects in the effective anisotropic shear; however, when we compare to observations (see below) we find no degeneracies between these two parameters. As the scale factor tends to zero, so it does $g(a,k)$ and it does not spoil CMB acoustic oscillations prior to last scattering.  It will, on the other hand, have effects on the ISW and the clustering properties of  matter fields.  For $g_0=0$ one recovers the case without anisotropic stress, $c_g=0$ implies scale free dependence, and for high $k$ the stress goes to zero, so it has no impact over small structures.

\section{Effects of DE anisotropic stress on power spectra} \label{results}

We adapted the codes \verb|CAMB|\footnote{\href{https://camb.info/}{https://camb.info/}} \cite{camb} and \verb|CosmoMC|\footnote{\href{https://cosmologist.info/cosmomc/}{https://cosmologist.info/cosmomc/}.} \cite{Lewis:2002ah} to include the shear contribution as detailed in the previous sections. 
We analyze the outcomes of the above anisotropic stress phenomenological model in combination with the effects of different DE EoS. The cosmological data set used in this work is:  BAO measurements from
6dFGS, SDSS-MGS, and BOSS LOWZ BAO \cite{Blakewigglez,10.1111/j.1365-2966.2011.19250.x6dFGS, 10.1093/mnras/stv154MGS, 10.1093/mnras/stw1264Boss},  supernovae 
from the combined Pantheon 
Sample \cite{Scolnic:2017caz}, recent $H_0$ measurement from Riess 2018 \cite{Riess_2018},  high-$\ell$ CMB TT spectrum and  low-$\ell$ polarization data from Planck 2015 \cite{2016A&A...594A..13P}. The main results are separated considering different DE EoS parametrizations that are detailed in the upcoming subsections.

\subsection{$w \approx -1$ }

\begin{figure}[htbp]
       \centering
      \includegraphics[width=.48\textwidth]{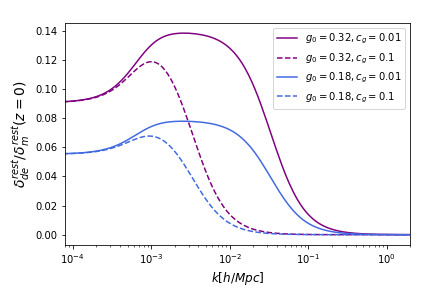}
       \caption{
       Ratios of DE to DM rest-frame densities at redshift $z=0$. We employ an EoS with $w_0=-1$ and various anisotropic stress parameters ($g_0, c_g$).
       }
       \label{fig:de-de/dm}
     \end{figure}
\begin{figure}[]
      \centering
      \includegraphics[width=.38\textwidth]{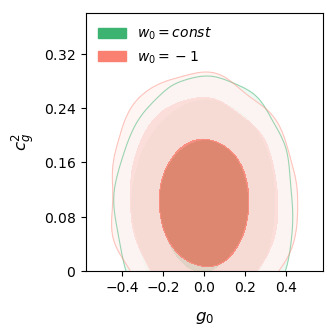}
       \caption{Contour confidence plots of the DE anisotropic parameters 
       $(g_0, c_g)$ at 68, 95 and 99\% c.l. for the $w = - 1$ and $w_0=\text{constant}$  
       EoS models. }
       \label{fig:2Dg0cgLCDM}
     \end{figure}

DE emulating a cosmological constant at background level  has no density perturbations in the absence of DE anisotropic stress.
However, one expects that evolving DE will have differences, albeit small, from $w=-1$. But, for all practical purposes many DE/MG models are indistinguishable from $\Lambda$CDM at background level, so for definiteness we adopt $w=-1$ in this subsection.
The inclusion of anisotropic stress generates fluctuations  that depend on the chosen anisotropic stress parameters: 
bigger $g_0$ values generate bigger perturbation amplitudes; and,  bigger $c_g$ values shift the anisotropic shear effects to larger scales.   These behaviors are shown in  Fig.~\ref{fig:de-de/dm}, where  ratios of DE  to matter rest-frame densities are plotted for parameters $g_0=0.18,\,0.32$ and $c_g=0.01,\,0.1$. These values are chosen because they lie inside the 1-$\sigma$  and 2-$\sigma$ confidence interval levels (c.l.) allowed by the Monte Carlo Makov Chain (MCMC) (as we will show in Fig.~\ref{fig:2Dg0cgLCDM}), and still provide large deviations to the matter PS, reaching a maximum of 15\% when compared to the $\Lambda$CDM ($g_0=0$) case, as explained below.

\begin{figure}[htbp]
       \centering
      \includegraphics[width=.48\textwidth]{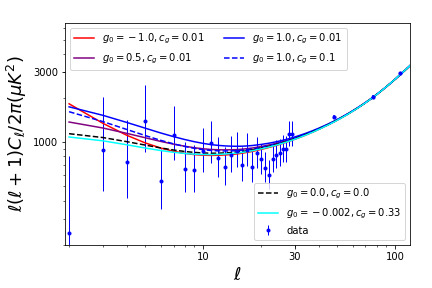}
       \caption{CMB TT power spectra at low multipoles for different $c_g$ and $g_0$ values and EoS $w_0 = - 1$. 
       The best fit values are $g_0=-0.0020, c_g=0.33$, as shown in Table \ref{tabla:best}.}
       \label{fig:ppfg0cg}
     \end{figure}

We varied the parameters ($g_0$, $c_g$) using \verb|CAMB| to obtain various CMB TT power spectra, as shown in Fig. \ref{fig:ppfg0cg}, along with CMB Planck data  \cite{2016A&A...594A..13P}. For $c_g$ fixed, positive $g_0$ values enhance low-$\ell$ anisotropies due to the ISW, whereas negative values diminish them, except for very low multipoles, where multipoles can go crossing the $\Lambda$CDM curve to overtake it.  
We also show the $\Lambda$CDM best fit (black dashed line), labeled in the figure as $g_0=0, c_g=0$.
We note that setting $g_0$ fixed, the effect of increasing $c_g$ is both to decrease the low-$\ell$ anisotropies and to shift their effect to smaller $\ell$-modes. This plot exhibits that, as we expected, the anisotropic parameters are not affecting high $\ell$-multipoles, where all the plotted curves coincide. For low-multipoles anisotropies, one may try to adjust downwards the curve, however, the relevance of these data points lies in the cosmic variance. In fact, low multipoles, including the quadrupole, turn out to be consistent with the $\Lambda$CDM model \cite{Bennett:2010jb}.   

\begin{figure}[htbp]
       \centering
      \includegraphics[width=.48\textwidth]{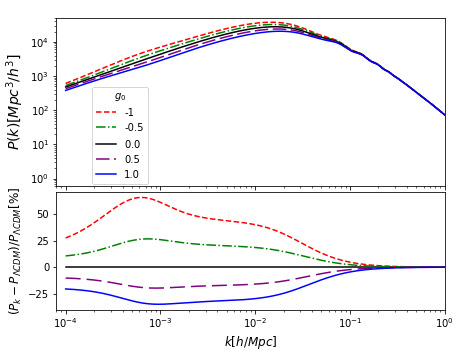}
       \caption{
       Matter PS produced by different $g_0$ values with $c_g=0.01$ fixed. The bottom panel shows the percent differences with respect to $\Lambda$CDM model. The background model is $w_0=-1$ EoS.
       }
       \label{fig:mpsppf2}
     \end{figure}

Large scale anisotropic stress imprints an effect on the clustering of matter at late times, as in the PS and growth function \cite{Koivisto:2005mm,Hu:1998kj,Mota:2007sz,Pogosian_2010, Sapone_2012,Cardona_2014,PhysRevD.90.103528}.  To see this, we plot the PS in Fig.~\ref{fig:mpsppf2} for 
($g_0$, $c_g$) parameter values such that $c_g$ is fixed and $g_0$ takes values between $[-1, 1]$, in a similar way we did it in Fig.~\ref{fig:ppfg0cg}.  
The effects over this matter statistic are clear: negative values of  $g_0$ tend to rise the PS for low $k$-modes, and for large $k$ we recover the $\Lambda$CDM model since $g(k\rightarrow 0) \rightarrow 0$;  positive $g_0$ values produce the opposite effect. Note that we have fixed the normalization such that all models have the same primordial power spectrum amplitude $A_s$ and spectral index $n_s$. For that reason all models coincide for modes $k_H \gg c_g^{-1}$, where $g$ becomes negligible. 

\begin{figure}[htbp]
       \centering
      \includegraphics[width=.48\textwidth]{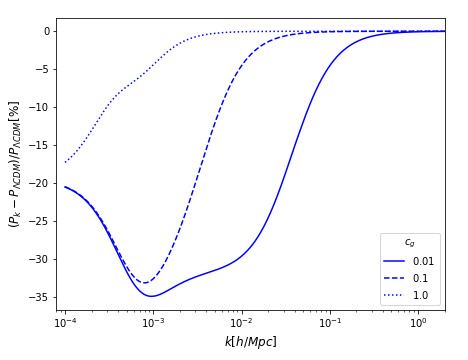}
       \caption{Percent deviations on the  PS relative to the $\Lambda$CDM case produced by different $c_g$ values with fixed $g_0=1$. The background model is $w_0=-1$ EoS.}
       \label{fig:mpsppfcg2}
     \end{figure}

Now, selecting some of the parameters of Fig.~\ref{fig:ppfg0cg}, we show in Fig.~\ref{fig:mpsppfcg2} again the percentage departures with respect to the $\Lambda$CDM model: $c_g \rightarrow 0$ increases the effect. These deviations in the PS amplitude can be large in the range of linear perturbations, and in fact they will also contribute to the nonlinear PS, that we shall study in Section \ref{nonlinear_PS}.        
     

\begin{figure}[]
       \centering
      \includegraphics[width=.48\textwidth]{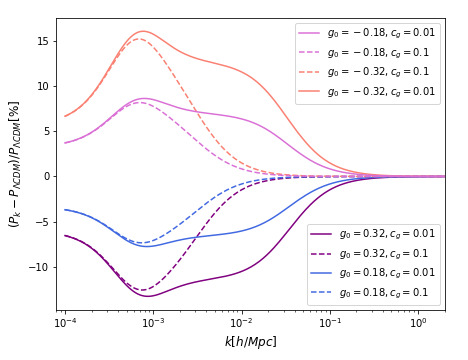}
       \includegraphics[width=.48\textwidth]{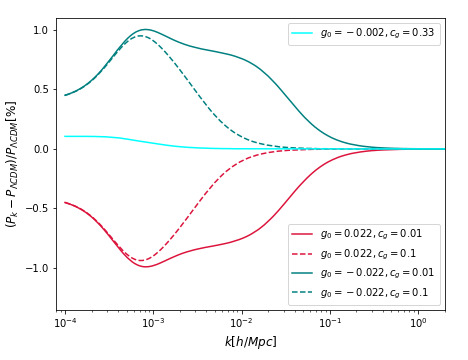}
       \caption{PS percentage deviation from $\Lambda$CDM. The anisotropic parameters ($ g_0$, $c_g$) considered are such that the maximum percentage deviation is of around 15\% (left panel) and of 1\% (right panel), corresponding to 4.09\% and to 0.28\% at a scale $k=0.05 \, h/\text{Mpc}$, respectively. We include the best fitted plot of our data set (see beginning of Section \ref{results}) corresponding to $g_0=-0.0020, c_g=0.33$ that produce a difference of around 0.05\% from $\Lambda$CDM model. These results are for the $w_0=-1$ EoS model.}
       \label{fig:percdif1-10}
     \end{figure}

We know, however, that deviations from the $\Lambda$CDM PS at scales $\sim 0.01$-$0.1 \, h/\text{Mpc}$ could not be as large as in Fig.~\ref{fig:mpsppf2} or \ref{fig:mpsppfcg2}, since these would affect the BAO features, and given the upcoming galaxy surveys such as DESI \cite{Aghamousa:2016zmz}, the constraints will tighten to uncertainties to be less than (or order of) $1\%$.  Consequently, we explore for deviations that are of the order of $1\%$  and at most $15\%$ 
(up to an overall normalization) in Fig.~\ref{fig:percdif1-10}, left and right panels, respectively. 
Note that the deviations from  $\Lambda$CDM reach their maximum around  $k\sim10^{-3} \, h/\text{Mpc}$. At the scale $k \sim 0.01 \, h/\text{Mpc}$, deviations are of 10\% (left panel, models $g_0=|0.32|$, $c_g=0.01$) and of 0.66\% (right panel, models $g_0=|0.022|, c_g=0.01$) and at $k\sim 0.05\, h/\text{Mpc}$ of 4\% (left panel) and of 0.3\% (right panel).

In both panels of Fig.~\ref{fig:percdif1-10}, dashed lines are for $c_g=0.1$, solid lines for $c_g=0.01$ and color changes for different $g_0$ as it is shown in the labels. Negative values of $g$ will increase the potentials wells, as can be seen from Eq.~(\ref{eq:isw}), and then the PS increases as well. Increasing $c_g$ lowers the absolute value of the maxima ($g_0 <0$) or minina ($g_0 >0$), but this is a by-effect of the produced shift along  $k$, that erases shear fluctuations above $k_H \sim c_g^{-1}$.  
Here we appreciate that deviations of at most 15\% are possible if $|g_0|< 0.32$.  In the right panel, 
we find $g_0$ values  in the interval $|g_0|\leq 0.022$ permit at most order $1\%$ deviations from the $\Lambda$CDM model. 

To find out what anisotropic parameter values are more realistic and preferred by cosmological data we perform an MCMC sampling of the parameter space using \verb|CosmoMC|  and the cosmological data set described at the beginning of this section. 
The relevant best-fit values for the $w=-1$ model without and with DE anisotropic stress are in columns I and II, respectively, of Table \ref{tabla:best}.   For anisotropic parameters we obtain $g_0=-0.0020^{+0.1474}_{-0.1523}, c_g^2=0.110^{+0.051}_{-0.069}$ at $68\%$ c.l.,  in agreement with the results of the vanilla $\Lambda$CDM cosmological parameters from Planck \cite{2016A&A...594A..13P}.

The CMB TT power spectrum produced by these values is included in Fig.~\ref{fig:ppfg0cg} (cyan color)
 that is alike the $\Lambda$CDM model, meanwhile other models vary only at low-multipoles, as already explained.
Similarly, in the right panel of Fig.~\ref{fig:percdif1-10}, we include the PS of our best-fit to obtain deviations of around 0.05\% at $k< 10^{-3} h/$Mpc from the no anisotropic stress case.

On the other hand, it is common that a comparison of models can be done using the Akaike information criterion (AIC) \cite{AIC} or the  Bayesian information criterion (BIC) \cite{BIC}, such that the model with the minimum AIC/BIC value is considered the best fit model. We compute the differences  $\Delta \textrm{AIC/BIC}$ with respect to $\Lambda$CDM, and the results are displayed in table \ref{tabla:best}.  For the $w=-1$ model the $\Delta \textrm{AIC}$ is considerable less supported than $\Lambda$CDM, and similarly $\Delta \textrm{BIC}$ shows that the $\Lambda$CDM model is stronger supported \cite{ref1, Kass:1995loi}. The other models in this work also show similar statistics. However, models with DE anisotropic stress possess more degrees of freedom and hence these statistics are a measure of how the added complexity turns into a less favored model. Our intention, though, was not to propose a more simple alternative to $\Lambda$CDM, but in fact to add anisotropic stress as a real possibility to the DE nature.

We finalize this subsection presenting the contour confidence region of the stress parameters in Fig.\ref{fig:2Dg0cgLCDM},  
confirming that the no-anisotropic case ($g_0 = 0$) is allowed at 1-$\sigma$ by CMB data. Nevertheless, the left panel of Fig.~\ref{fig:percdif1-10} shows that for anisotropic parameter values that are inside the 2-$\sigma$ best-fit values, they produce differences on the PS 
of at least 15\% with respect to $\Lambda$CDM. In this parameter range the CMB will be well fitted and not changing significantly. It is then clear that the PS imposes tighter constraints than the CMB on the anisotropic stress and hence  is a potential theory discriminator of different DE anisotropic stress models.

\begin{table*}[width=.99\linewidth,t]
\scriptsize
\centering
\caption{Best-fit values and marginalized 0.68 confidence intervals for our cosmological data set Pantheon, BAO, CMB TT and low-P. Column I corresponds to $\Lambda$CDM model, column II to $w=-1$ EoS with DE anisotropic stress allowed, in column III the parameter $w_0$ is a free constant, in column IV the CPL parameterization with $w_0, w_a$ as free parameters, in column V the 7CPL parameterization with no anisotropic stress, and in column VI  the 7CPL parameterization with anisotropic stress. At the bottom of the table we include the values of two model comparison criteria (AIC/BIC) with respect to $\Lambda$CDM model. 
}
\begin{tabular}{lcccccc}
\hline
& & & &\\[-3pt]
$\qquad$ &\phantom{ab} $\qquad$ I$\qquad$\phantom{ab}  & \phantom{ab}$\qquad$ II $\qquad$  & $\qquad$ III $\qquad$ & $\qquad$ IV $\qquad$ &$\qquad$ V $\qquad$    &$\qquad$ VI $\qquad$  \\ [2pt] 
$\qquad$  & $\Lambda$CDM & $w=-1$ & $w_0$ constant & CPL  & 7 CPL no-stress & 7 CPL \\[5pt]
\hline 
& & & & & \\[1pt]
$\Omega_{b}h^2$ &  $0.0224\pm 0.0002$ & $0.0224\pm 0.0002$ & $0.0224\pm 0.0002$ & $0.0222\pm 0.0002$ & $0.0223\pm 0.0002$ & $0.0223\pm 0.0002$\\[5pt]
$\Omega_c h^2$& $0.1179\pm 0.0012$&$ 0.1179\pm 0.0012$ & $0.1190\pm 0.0016$ & $0.1199\pm 0.0019$ & $0.1192\pm 0.0016$ & $ 0.1192\pm 0.0015$\\[6pt]
$\tau$ & $0.086\pm 0.018$ & $ 0.085\pm 0.018$ & $0.081\pm 0.018$ & $ 0.077\pm 0.019$ & $0.080\pm 0.019$ &$0.080\pm 0.019$\\[6pt]
${\rm ln}(10^{10}A_{s})$& $3.102\pm 0.036$ & $3.101\pm 0.037$ & $3.095\pm 0.036$ & $3.088\pm 0.037$ & $3.094\pm 0.036$ & $3.092\pm 0.037$\\[6pt]
$n_s$&  $0.9699\pm 0.0043$ & $0.9698\pm 0.0044$ & $0.9673\pm 0.0051$ & $0.9654\pm 0.0055$ & $ 0.9670\pm 0.0051$ & $0.9669\pm 0.0049$\\[5pt]
$H_0$& $68.17\pm 0.52$ & $ 68.16\pm 0.52$ & $68.73\pm 0.81$ & $68.74\pm 0.80$ & $68.73\pm 0.80$ & $68.78\pm 0.81$\\[6pt]
$w_0$& $-1.00$ &$-1.00$& $-1.036\pm 0.037$ & $-0.977\pm 0.085$ & $-1.031\pm 0.037$ & $-1.032\pm 0.037$\\[6pt]
$w_a$& $0$ &$0$& 0 & $-0.29^{+0.42}_{-0.31}$ & $<0.818$ & $<0.882$ \\[6pt]
$g_0$& $0$ & $-0.0020 ^{+0.1474}_{-0.1523}$& $-0.0008^{+0.1509}_{-0.1521}$ & $-0.0004^{+0.1508}_{-0.1545}$ & $0$& $-0.0026^{+0.1535}_{-0.1529}$\\[6pt]
$c_g^{2}$ & $0$ &$0.110^{+0.051}_{-0.069}$ & $0.108^{+0.051}_{-0.069}$& $ < 0.070$ & $0$& $< 0.070$ \\[6pt]
\hline
& & & & & \\[1pt]
$\Delta$AIC & $0$ &$5.95$ & $7.82$& $10.52$ & $3.84$ & $9.78$ \\[6pt]
$\Delta$BIC & $0$  &$18.35$ & $26.43$& $35.36$ & $16.25$ &$34.59$ \\[6pt]
\hline
\end{tabular}
\label{tabla:best}
\end{table*}

\subsection{{\it w =} constant}
In this section the expansion history is slightly different from $\Lambda$CDM, now we assume $w=w_{0}$ constant. Thus, density fluctuations are generated even if DE does not possess anisotropic stress,  
showing that perturbations attenuate as $w \rightarrow -1$ and when $c_{s} \rightarrow 1 $.

\begin{figure}[]
    \centering
        \includegraphics[width=.48\textwidth]{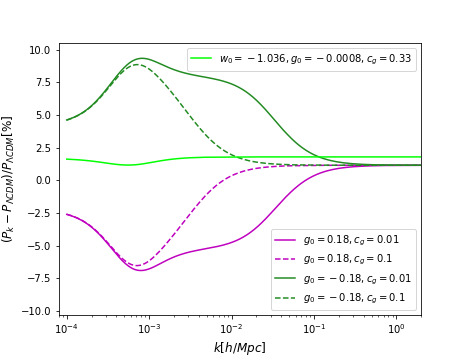}
    \caption{ 
    PS percentage deviation with respect $\Lambda$CDM.  
    In these plots $w_0=-1.023$, and $g_0$, $c_g$ are as the labels indicate; we include the best-fit curve from Table \ref{tabla:best}.}
    \label{fig:cmbw0g0}
\end{figure}

For this model, anisotropic stress parameters leave very similar imprints on the CMB TT curve as those of the $w=-1$ model, see Fig. \ref{fig:ppfg0cg}, so we omit to show these results, and the same discussion about the  parameters ($g_0,\, c_g$) prevails for this EoS.  The effects on the PS  are shown in Fig.\ref{fig:cmbw0g0}, where the EoS reference value is $w_0=-1.023$ in agreement with Planck's results ~\cite{2016A&A...594A..13P} and the values $c_g$ and  $g_0$ were chosen so that they result in visible changes, with deviations from $\Lambda$CDM of order of $10\%$ or less.     
For these plots we obtain a maximum deviation of around 10\% at $k\sim 10^{-3} h/\text{Mpc}$, and of  $3.4\%$ at $k = 0.05 h/\text{Mpc}$.
All anisotropic stress values we used to generate  Fig.~\ref{fig:cmbw0g0} 
are consistent with the Planck CMB TT  measurements.  

\begin{figure}[htbp]
    \centering
    \includegraphics[width=.38\textwidth]{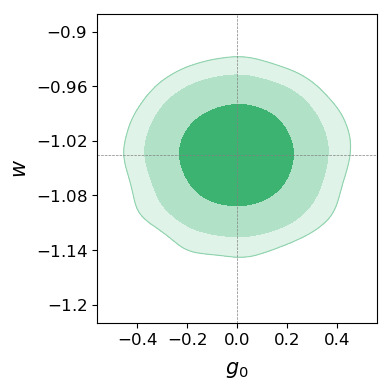}
    \includegraphics[width=.38\textwidth]{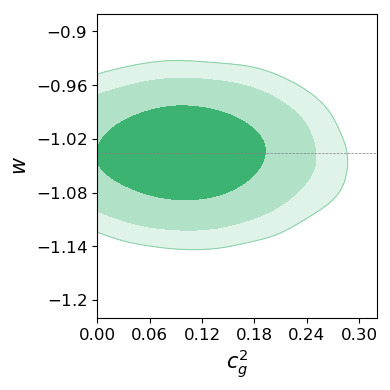}
    \caption{Contour confidence plots at $68$, $95$ and $99\%$ for the $w_0=$constant 
     EoS, corresponding to results in  column III of Table \ref{tabla:best}.  
    }
    \label{fig:contourwCDM}
\end{figure}

The relevant results of an MCMC fit are presented in column II  of Table \ref{tabla:best}, and the contour plots that involve DE parameters are shown in Figs.~\ref{fig:2Dg0cgLCDM} and \ref{fig:contourwCDM}. The resulting parameters $g_0$ and $c_g^2$ are similar to the ones produced by the $w_0=-1$ EoS.
As expected, the values $w_0=-1, g_0=0$ are inside the 68\% contours, meaning that the results are in agreement with the $\Lambda$CDM model.

Finally, we note that the values of $c_g$, $g_0$ reported in Table \ref{tabla:best} are similar to the other models. This motivated us to show in Fig.~\ref{fig:contourwCDM} the contour plots of the anisotropic parameters with the EoS parameter to clarify any degeneracy among them. We found  essentially no degeneracy in $w_0$ and $g_0$, and a small effect in $w_0$ and $c_g$, as also proved in their  corresponding correlation matrices in a principal component analysis.
\subsection{Thawing parametrization (CPL)} 
Now we consider the CPL EoS, 
providing a thawing behavior for $w_a < 0$. CPL is one of the most popular EoS for DE, and according to Planck 2015
results \cite{2016A&A...594A..13P} in the absence of DE anisotropic stress, the best fit values 
for the data set we are using are  $w_0=-0.93^{+0.23}_{-0.22}$, $w_a=-0.41^{+0.87}_{-0.91}$ at 2-$\sigma$ \cite{planck_tab}, that we will take as reference values. 

\begin{figure}[ht]
    \centering
    \includegraphics[width=.48\textwidth]{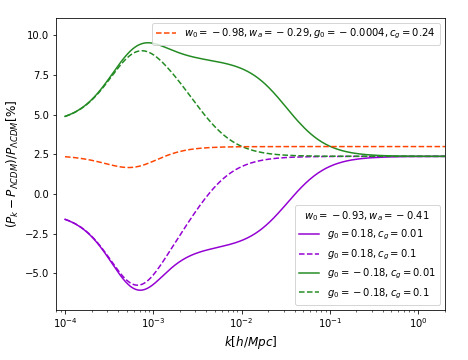}
    \caption{Deviations on the PS, produced by CPL parameterization and DE anisotropic stress. The parameters of DE  EoS are $w_0=-0.93$, $w_a=-0.41$, that are the Planck's reference values. We also include our best DE fitted parameters, shown in column III of Table \ref{tabla:best}. }
    \label{fig:MPScpl}%
\end{figure}

We found similar effects due to the anisotropic parameters on the CMB TT power spectrum for this EoS. We plot in Fig.~\ref{fig:MPScpl} our results on deviations of the PS with respect to $\Lambda$CDM model, 
as in Fig.~\ref{fig:percdif1-10} (left panel) and Fig.\ref{fig:cmbw0g0}, yielding maximum differences of around $10\%$; particularly, at $k = 0.05 \, h/\text{Mpc}$ the maximum deviation is about 4\%. 

\begin{figure}[htbp]
       \centering
      \includegraphics[width=.38\textwidth]{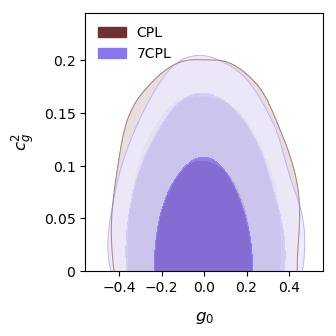}
       \caption{
       Contour confidence plots at 68, 95, and 99\% of the parameters  $(g_0, c_g)$ for models CPL and 7CPL.
       }
       \label{fig:confidenceg0cg_nCPL}
     \end{figure}

Finally, we performed
an MCMC statistical analysis parametrizing $w_{de}$ as CPL. The best-fit parameters and c.l. at 68\% are presented in column III of Table \ref{tabla:best}, and their corresponding  contour plots for $(g_0, c_g)$ in Fig.~\ref{fig:confidenceg0cg_nCPL}. The contour plot for anisotropic stress parameters looks very similar to the one obtained with the EoS $w_0=-1$; see Fig.~\ref{fig:2Dg0cgLCDM}, which indicates that the DE EoS and stress parameters are quite independent.

\subsection{Freezing parametrization (7-CPL)}
\begin{figure}[]
       \centering
      \includegraphics[width=.38\textwidth]{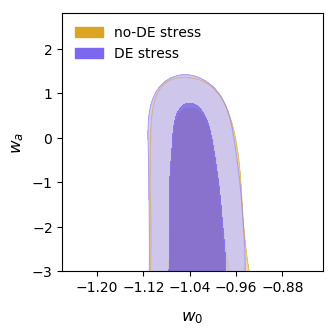}
       \caption{Contour confidence plots at 68, 95, and 99\% for 7CPL EoS parameters with and without DE anisotropic stress.
       }
       \label{fig:confidencew0wancpl}
     \end{figure}

Finally we consider the n-CPL DE parametrization, Eq. (\ref{eq:ncpl}) with $n=7$ which has a freezing behavior if $w_0<0$ and $w_a>0$. To our knowledge, for this EoS there are no reported best-fitted values for $w_0, w_a$. Then, we first estimate them for the case of null anisotropic stress. The results are shown in column IV of Table \ref{tabla:best} and the contour plot is presented in Fig.\ref{fig:confidencew0wancpl} (yellow regions). 
     For this EoS the standard cosmological model is recovered ($w_0=-1$, $w_a = 0$) at 68\%. 
     The $w_0$ parameter is well restricted by late time observations, whereas $w_a$ is not sensitive to these cosmological data set, its upper limit at 68\% is 0.882 but it can take a wide range of negative values. 
 This is consistent with claims in the sense that fittings suggest thawing \cite{Planck2016,Hara:2017ekj} and freezing models \cite{PhysRevD.93.103503}; we find that both are allowed for this EoS. 
Variations on $w_a$ are not 
visible in the CMB TT, but deviations are present in the amplitude of the matter PS. When varying $w_a$ from $-1.1$ to $0.8$ the change in the PS is of about 1$\%$ in linear scales and these tend to lower the power, the larger (positive) $w_a$ values are.  In this case, the effects in the PS occur in $k$ between  $10^ {-4}$ and $10^{-3}$ $h/$Mpc; after $k =0.004 \, h$/Mpc the PS behaves as $\Lambda$CDM. 
     
Now we introduce DE anisotropic stress as in the above subsections. The data fits are shown in column V of Table \ref{tabla:best},  and their contour plots concerning to DE stress parameters are in Fig.~\ref{fig:confidenceg0cg_nCPL}. 
The contour plot $w_a-w_0$ is shown in Fig.~\ref{fig:confidencew0wancpl} (purple regions) together with that of no-anisotropic stress. The similarity between both cases shows that
data from CMB and late time background evolution are agnostic to the presence of the DE anisotropic stress.

\begin{figure}[htbp]
       \centering
      \includegraphics[width=.48\textwidth]{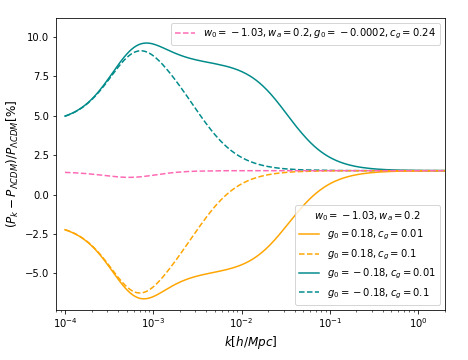}
       \caption{Deviations on the PS of around $10\%$ caused by different anisotropic stress values in the 7CPL parametrization, with freezing parameters $w_0=-1.03, w_a=0.2$. We also include the best fitted shear parameters curve.
       }
       \label{fig:MPS7cplwag010}
     \end{figure}

In Fig.\ref{fig:MPS7cplwag010} we include the best-fitted values of this model and others inside the 95\% contour confidence plot to obtain deviations on the PS of around $10\%$ with respect $\Lambda$CDM. 

\bigskip 

The lesson from all these models is that they leave particular and potentially detectable features in the PS, even when the parameter space for DE EoS and anisotropic parameters are allowed by CMB and  background probes.  In general, we found that the anisotropic stress provokes deviations smaller than $10\%$ with respect to the $\Lambda$CDM PS at  $k \sim 0.01 \,h/\text{Mpc}$ for the parameters in the range  $-0.30< g_0 < 0.32$, $0 \leq c_g^2 < 0.01$ and smaller than $5\%$  for $-0.15 < g_0 < 0.16$, $0 \leq c_g^2 < 0.01$.


\section{Nonlinear evolution}\label{nonlinear_PS}
Our treatment until here has been limited to linear physics. In the present section we now explore the PS behavior produced by the  nonlinear evolution at quasi-linear scales. That is, we will consider some selected values of the above parameters and evolve the system with the standard tools of nonlinear perturbation theory. We will consider the real space PS obtained using standard perturbation theory and the redshift-space multipoles using the TNS model \cite{Taruya_2010}. 
Hence, the objective of this section is to show how the features produced by the dark energy anisotropic stress in the linear power spectrum evolve when one considers the quasi-linear regime both in real and redshift space. 
Since such features are located in linear and mildly non-linear regions of the power spectrum, they are amenable for a standard study within perturbation theory, and it is natural to expect that they become enhanced when compared to the linear results. We remark that the results of this section are not used to compare to  observations; that would be erroneous, because one would has to tailor a halofit model for these specific models, which is beyond the scope of this work.

The most common approach adopted to study MG/DE models beyond the linear regime is by parametrizing the Poisson equation 
\cite{Zhao:2008bn} with a function $\mu(k,z)$, which quantifies the deviation of the gravitational Newton constant from the expected value in GR without DE perturbations; in DE models with anisotropic stress, this should be understood as due to the differences in the gravitational potentials $\Psi$ and $\Phi$, and not as a modification to GR, as discussed in Section \ref{DE-ani_stress}. The Poisson equation reads 
\begin{eqnarray}\label{eq:Poissonmu}
\frac{k^{2}}{a^2} \Phi 
= -4\pi G \mu(k,z) \rho \Delta, \end{eqnarray}
with $\Delta=\delta \rho^\text{rest}_m / \rho$ the matter density contrast. Since nonlinearities are developed at late time we can neglect the contributions of radiation components, and keep only those of matter (combined CDM + baryons) and DE. We identify
\begin{equation}\label{eq:mu}
\mu(k,z)=1+\frac{\delta \rho_{de}^{rest}}{\delta \rho_{m}^{rest}}.
\end{equation}
In our case, we can construct the function $\mu(k,z)$ with the linear theory developed in the previous sections and with the ingredients obtained from \verb|CAMB|. Notice that in writing Eq. (\ref{eq:Poissonmu}) we can approximate the matter gauge-invariant overdensity as $\Delta \approx \delta =\delta \rho /\rho$, since matter particles are non-relativistic and we are interested on scales well inside the Hubble length $H^{-1}$. With Eqs.~(\ref{eq:Poissonmu}) and (\ref{eq:mu}) we can construct a Standard Perturbation Theory (SPT) following the recipes developed in Ref.~\cite{Koyama_2009,Taruya:2016jdt,Aviles:2017aor} for MG models, that can be adapted to DE models with stress. The formalism consists 
first in constructing the kernels $F_n$ that appear in the higher than linear order overdensities, such that at order $n$ in SPT one has
\begin{align}
\delta^{(n)}(\vk) &= \int \frac{d^3p_1 \cdots d^3 p_n}{(2\pi)^{3(n-1)}}
F_n(\vp_1,\cdots,\vp_n)  \delta^{(1)}(\vp_1) \cdots \delta^{(1)}(\vp_n),
\end{align}
where $\delta^{(1)}$ is the linear matter overdensity treated in the previous sections. Thereafter, we can construct the corrections to the linear PS by computing the correlations $\langle \delta^{(2)}(\vk)\delta^{(2)}(\vk')\rangle$ and 
$\langle \delta^{(1)}(\vk)\delta^{(3)}(\vk')\rangle$, from which we obtain
the first, 1-loop, correction to the linear PS
\begin{equation}
P^\text{1-loop}(k) = P_L(k) + P_{22}(k) + P_{13}(k),    
\end{equation}
with
\begin{align}
  P_{22}(k) &=  2 \int \frac{d^3p}{(2\pi)^3} F_2(\vp,\vk-\vp) P_L(|\vk-\vp|)P_L(p), \\
  P_{13}(k) &= 6 P_L(k) \int \frac{d^3p}{(2\pi)^3} F_3(\vp,-\vp,\vk) P_L(p).
\end{align}
Notice that for primordial Gaussian distributed density fields, as we are considering here, correlations  $\langle \delta^{(n)}(\vk)\delta^{(m)}(\vk')\rangle$, with $n+m$ an odd number, vanish.

The main obstacle to obtain the nonlinear PS to 1-loop is to find the kernels
$F_n$. However, note from Fig.~\ref{fig:de-de/dm} that the we can safely approximate $\mu(k,z) \approx 1$ at quasi-linear and non-linear scales, where the error we are introducing on $\mu$ is a at most a few percent. $\mu$ departs from unity the most at low-$k$, hence not influencing significantly the mildly nonlinear scales $k\sim 0.1 \,h/\text{Mpc}$.   To have an idea when these deviations should be accounted for, consider $f(R)$ MG gravity where $\mu$ interpolates between 1 at large scales and $4/3$ at nonlinear high-$k$ values. And even in these cases the use of the well known Einstein-de Sitter $F_n$ kernels is not a bad approximation. If found necessary, one can reintroduce the scale dependence on the function $\mu$.

Having at hand the SPT kernels for density fields, and the corresponding $G_n$ kernels for velocity fields one can compute, apart from the loop corrections in real space, the loop corrections in redshift space. Here we will adopt the popular TNS model  to do this. 

As we have discussed, perhaps the most interesting models we have studied are those with $w=-1$, because we can observe the differences in the angular CMB spectrum and matter PS only due to the anisotropic stress and not to a different background evolution. Hence, here we focus on that model with anisotropic parameters $g_0=0.32$, $c_g=0.01$, whose linear
PS is shown in the left panel of Fig.~\ref{fig:percdif1-10}, and its overdensities ratio in Fig.~\ref{fig:de-de/dm}. From the models presented in Fig.~\ref{fig:percdif1-10}, those with $c_g=0.1$ are not very interesting from the point of view of SPT because the differences with the $\Lambda$CDM  PS lie at very large, still linear scales, and hence the nonlinear corrections are negligible; the model  $g_0=-0.32$, $c_g=0.01$ result in the same qualitative behavior but with an opposite sign, reflected upwards with respect to the $\Lambda$CDM PS (a similar effect seen in Fig.~\ref{fig:percdif1-10}). 
We choose $g_0=-0.32$ instead of the value $g_0=-0.18$, which is allowed by observations at 1-$\sigma$, in order to enhance the differences with the $\Lambda$CDM model.

\begin{figure}[]
       \centering
      \includegraphics[width=.48\textwidth]{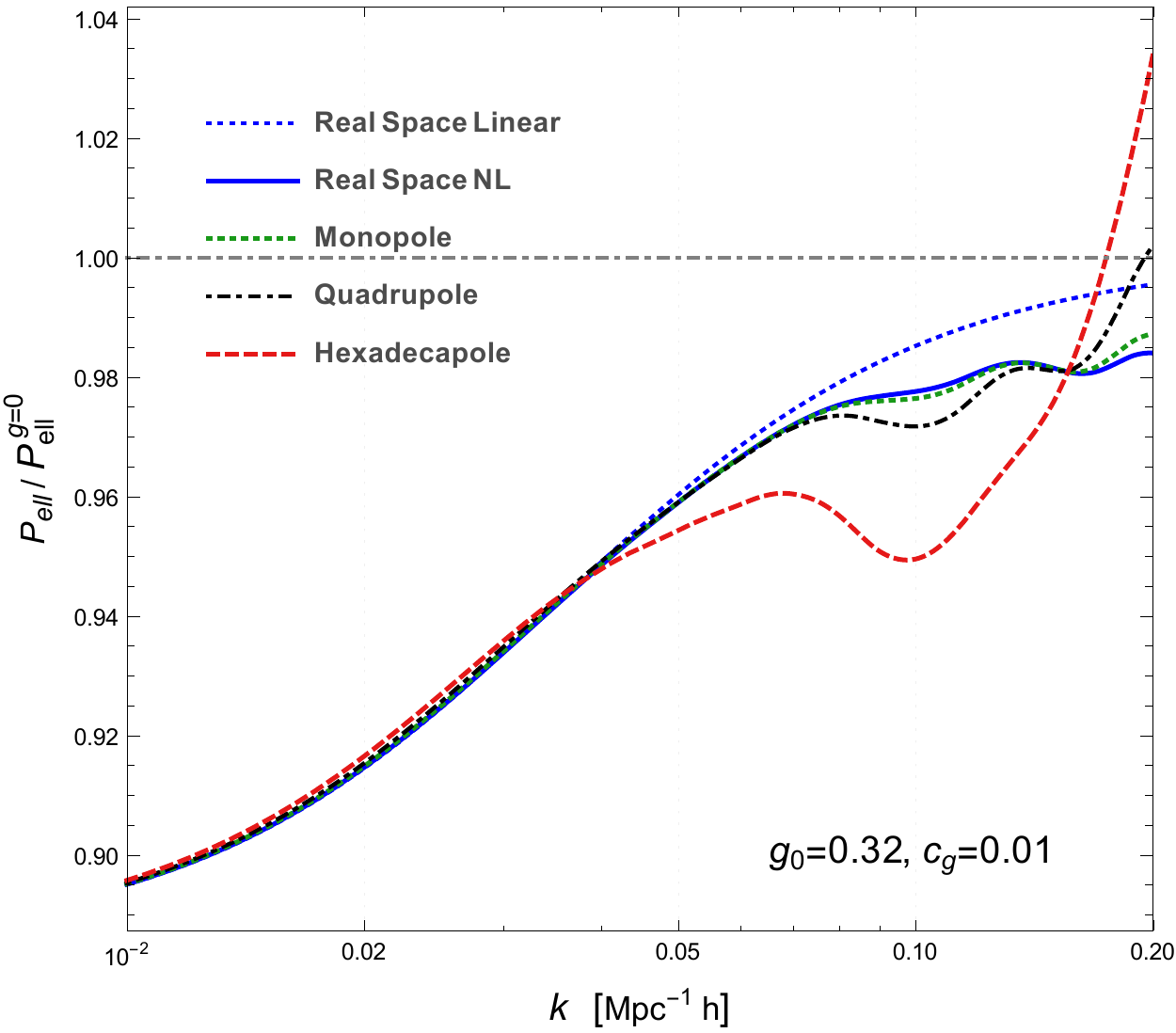}
       \caption{1-loop PS and RSD multipoles for the anisotropic stress model $g_0=-0.32$, $c_g=0.01$. The ratio to the $\Lambda$CDM is taken for comparison. The dotted blue line is for the linear real space case, already shown in the left panel of Fig.~\ref{fig:percdif1-10}. The solid blue line is for the real space 1-loop SPT PS. The rest  curves are the multipoles of the RSD TNS model: dotted green, monopole; dot-dashed black, quadrupole; and, dashed red, hexadecapole. }
       \label{fig:RSDmultipoles}
     \end{figure}

Figure \ref{fig:RSDmultipoles} shows our results divided by the corresponding power spectra in the $\Lambda$CDM to note more clearly the differences among the models. We have computed the 1-loop real space PS (solid blue line), and the monopole (dotted green), quadrupole (dot-dashed black), and hexadecapole (dashed red) multipoles of the 2-dimensional redshift space PS. Also, for comparison, we plot the linear real space PS (already shown in the left panel of Fig.~\ref{fig:percdif1-10}). 
The worthy point to note is that when considering  nonlinear evolution the differences with and without anisotropic stress are larger, and in redshift-space these are even more enhanced: 
the quadupole exhibits differences of the order of 3\% at $k \sim 0.1$ $h/$Mpc and
the hexadecapole of 5\%, the latter being the one that shows the largest departures from the $\Lambda$CDM  ---which unfortunately is also the one that presents a smaller signal-to-noise in observations and simulations.

\bigskip

\section{CONCLUSIONS} \label{conclusions}
The standard model of cosmology assumes that the recent accelerated expansion of the Universe is originated by a cosmological constant, but it may well be caused by an evolving piece, DE or MG. These two latter general schemes are degenerated at first order perturbation theory when DE is provided with anisotropic stress \cite{Kunz_2007,Song:2010,Arjona_2019,Arjona_2019b}. This is the reason that permits to introduce the DE stress in the context of the PPF formalism as an effective MG phenomenon, thus allowing to study its effects to the DE.  In general, DE perturbations are smaller than DM ones, but still they may leave an imprint on the CMB and clustering evolution. The role of anisotropic stress is to create (in the $w=-1$ EoS model) or amplify/modify DE density perturbations; other  effects were known to happen due to changes in the DE EoS or in its sound speed \cite{Weller:2003hw,dePutter:2010vy,Huterer_2017}.  Anisotropic stress has an impact in the ISW effect \cite{Hu:1998kj,Koivisto:2005mm,PhysRevD.77.103524,Ichiki_2007,Calabrese_2011,Chang_2014}, that results similar for the various EoS studied in this work.  But given the level of uncertainties due to the cosmic variance, CMB data will not shed light on such a component. However, current and forthcoming galaxy surveys can delimit this possibility.  

We studied the influence of anisotropic stress parameters using the PPF formalism \cite{PhysRevD.77.103524,PhysRevD.78.087303} in which we employed an ansatz on the anisotropic stress function with two parameters, one mainly controlling the amplitude ($g_0$) and the other the scale dependence ($c_{g}^{2}$), such that for early times or small scales, $k \gg a H c_g^{-1}$, the stress vanishes.  The best fitted parameters are shown in Table \ref{tabla:best} for the different EoS considered in this work.  All models predict that anisotropic stress parameters are consistent with $\Lambda$CDM model up to error bars. However, the possibility of nontrivial anisotropic stress is open.  Independent of the EoS parametrization, positive $g_0$ values make the perturbations to increase, the CMB TT low multipoles also increase, but the PS decreases with respect to the $\Lambda$CDM PS (negative $g_0$ values do the opposite). Further, we found that the parameters of the anisotropic stress and the EoS are not degenerated.

For the $w=-1$ model, CMB analysis allows any pair of values over the intervals $-1\leq g_0 \leq 1$, $0.01\leq c_g \leq 1$, but these are wide enough to 
produce large effects in the PS.  In fact, parameters in the range $0.5\leq |g_0|\leq 1$,  $0.01\leq c_g \leq 1$ reach differences with respect to $\Lambda$CDM of up to $30\%$,  which are too big to be acceptable.  The maximum percentage difference is driven by the $g_0$ value. In order for the anisotropic stress not to provoke deviations, with respect to $\Lambda$CDM, larger than $15\%$ in the PS, the $g_0$ parameter has to be in the range $|g_0|\leq 0.32$ and for deviations of up 1$\%$ the parameter should be in the range $|g_0|\leq 0.022$.  For the rest of the models considered in this work, $w$CDM, CPL, and 7CPL,  the deviations are similar in the parameter ranges just mentioned. In general for all models, we found that in order for the anisotropic stress not to provoke deviations larger than $10\%$ and $5\%$  with respect to the $\Lambda$CDM PS at $k \sim 0.01 \,h/\text{Mpc}$, the parameters have to be in the range  $-0.30< g_0 < 0.32$, $0 \leq c_g^2 < 0.01$ and $-0.15 < g_0 < 0.16$, $0 \leq c_g^2 < 0.01$, respectively.

We computed the PS at 1-loop using SPT and observe that the differences between models with and without anisotropic stress are amplified by  nonlinear evolution. We also obtained the 2-dimensional PS in redshift space using the TNS model and  show that these differences are even more enhanced, particularly for the quadrupole and hexadecapole multipoles.

Since one expects that present and future galaxy surveys will have uncertainties in the determination of the PS of  one-percentage levels, they could delimit  the anisotropic stress stemming from DE, or equivalently  from MG, to shed light on the nature of one of most mysterious components of the Universe. 

\section*{Acknowledgements}
The authors acknowledge partial support by CONACYT project 283151.
GG-A acknowledges CONACYT for grant no. 290778.

\printcredits

\bibliographystyle{cas-model2-names}

\bibliography{cas-refs}
\end{document}